# Exact and Approximation Algorithms for DNA Tag Set Design*

Ion I. Măndoiu and Dragoş Trincă

CSE Department, University of Connecticut
371 Fairfield Rd., Unit 2155, Storrs, CT 06269-2155
{ion.mandoiu,dragos.trinca}@uconn.edu

**Abstract.** In this paper we propose new solution methods for designing tag sets for use in universal DNA arrays. First, we give integer linear programming formulations for two previous formalizations of the tag set design problem, and show that these formulations can be solved to optimality for instance sizes of practical interest by using general purpose optimization packages. Second, we note the benefits of periodic tags, and establish an interesting connection between the tag design problem and the problem of packing the maximum number of vertex-disjoint directed cycles in a given graph. We show that combining a simple greedy cycle packing algorithm with a previously proposed alphabetic tree search strategy yields an increase of over 40% in the number of tags compared to previous methods.

## 1 Introduction

Recently developed *universal DNA tag arrays* [7, 13, 15] offer a flexible and cost-effective alternative to custom-designed DNA arrays for performing a wide range of genomic analyses. A universal tag array consists of a set of DNA strings called *tags*, designed such that each tag hybridizes strongly to its own *antitag* (Watson-Crick complement), but to no other antitag. A typical assay based on universal tag arrays performs Single Nucleotide Polymorphism (SNP) genotyping using the following steps [4, 9]: (1) A set of *reporter oligonucleotide probes* is synthesized by ligating antitags to the 5' end of primers complementing the genomic sequence immediately preceding the SNP. (2) Reporter probes are hybridized in solution with the genomic DNA under study. (3) Hybridization of the primer part (3' end) of a reporter probe is detected by a single-base extension reaction using the polymerase enzyme and dideoxynucleotides fluorescently labeled with 4 different dyes. (4) Reporter probes are separated from the template DNA and hybridized to the universal array. (5) Finally, fluorescence levels are used to determine which primers have been extended and learn the identity of the extending dideoxynucleotides.

Tag set design involves balancing two conflicting requirements: on one hand we would like a large number of tags to allow assaying a large number of biochemical reactions, on the other hand we would like the tags to work well for a wide range of assay types and experimental conditions.

Ben Dor et al. [3] have previously formalized the problem by imposing constraints on antitag-to-tag hybridization specificity under a hybridization model based on the classical 2-4 rule, and have proposed near-optimal heuristics. In Section 3 we give an integer linear programming (ILP) formulation for this problem and its variant in which tags are required to have equal length [14]. Empirical results in Section 5 show that these ILP formulations have extremely small integrality gap, and can be solved to optimality for instance sizes of practical interest by using general purpose optimization packages.

Previous works on tag set design [3, 14] require for substrings that may form a nucleation complex and initiate cross hybridization not to be repeated within any selected tag. This constraint simplifies analysis, but is *not* required for ensuring correct tag functionality – what is required is for such substrings not to appear simultaneously in two different tags. To our knowledge, no previous work has assessed the impact that adding this constraint has on tag set size. In this paper we propose two algorithms for designing tag sets while relaxing this constraint. The first one is a modification of the alphabetic tree search strategy in [13, 14], The second algorithm stems from the observation that periodic tags, particularly those with a short period, use the least amount of "resources" and lead to larger tag sets, where the limited resources are in this case minimal substrings that can form nucleation complexes (for formal models see Section 2). In Section 4 we establish an interesting connection between the tag design problem and the problem of packing the maximum number of vertex-disjoint directed cycles in a given graph, and propose a simple greedy algorithm for the latter one. Results in Section 5 show that combining the greedy cycle packing algorithm with alphabetic tree search strategy yields an increase of over 40% in the number of tags compared to previous methods.

---

* Work supported in part by a Large Grant from the University of Connecticut's Research Foundation.

The rest of the paper is organized as follows. In Section 2, we describe the hybridization model, give formal problem formulations for several variants of the tag set design problem, and briefly review previous work. In Section 3, we give integer program formulations for the problem variant in [3] and its variant in which tags are required to have equal length [14]. In Section 4, we describe the algorithms for problem variants in which potential nucleation complexes can be appear multiple times as substrings of a tag, and prove that packing the maximum number of vertex-disjoint directed cycles in a given graph is APX-hard. Finally, we give experimental results in Section 5 and conclude with some open problems in Section 6.

## 2   Problem Formulations and Previous Work

A main objective of universal array designers is to maximize the number of tags, which directly determines the number of reactions that can multiplexed using a single array. At the same time, tag sets must satisfy a number of *stability* and *non-interaction* constraints [6]. The set of constraints depends on factors such as the array manufacturing technology and the intended application. In this section we formalize the most important stability and non-interaction constraints using the hybridization model in [3].

**Hybridization model.** Hybridization affinity between two oligonucleotides is commonly characterized using the *melting temperature*, defined as the temperature at which half of the duplexes are in hybridized state and the other half are in melted state. However, accurate melting temperature estimation is computationally expensive, e.g., estimating the melting temperature between two non-complementary oligonucleotides using the near-neighbor model of SantaLucia [17] is an NP-hard problem [10]. Ben-Dor et al. [3, 4] formalized a conservative hybridization model based on the observation that stable hybridization requires the formation of an initial *nucleation complex* between two perfectly complementary substrings of the two oligonucleotides. For nucleation complexes, hybridization affinity is modeled using the classical *2-4 rule* [18], which estimates the melting temperature of the duplex formed by an oligonucleotide with its complement as the sum between the number of *weak* bases (i.e., A and T) and twice the number of *strong* bases (i.e., G and C).

The *weight* $w(x)$ of a DNA string $x = a_1 a_2 \ldots a_k$ is defined as $w(x) = \sum_{i=1}^{k} w(a_i)$, where $w(\texttt{A}) = w(\texttt{T}) = 1$ and $w(\texttt{C}) = w(\texttt{G}) = 2$. Throughout this paper we assume the following *c-token hybridization model* [3]: hybridization between two oligonucleotides takes place only if one oligo contains as substring the complement of a substring of weight $c$ or more of the other, where $c$ is a given constant. The *complement* of a string $x = a_1 a_2 \ldots a_k$ over the DNA alphabet $\{\texttt{A}, \texttt{C}, \texttt{T}, \texttt{G}\}$ is defined as $\bar{x} = b_1 b_2 \ldots b_k$, where $b_i$ is the Watson-Crick complement of $a_{k-i+1}$.

**Hybridization stability.** Current industry designs require a predetermined tag length $l$, e.g., GenFlex tag arrays manufactured by Affymetrix use $l = 20$ [1]. The model proposed in [3] allows tags of unequal length and instead require a minimum tag weight of $h$, for a given constant $h$. In this paper we consider both types of stability constraints, and use the parameter $\alpha \in \{l, h\}$ to denote the specific model used for hybridization stability.

**Pairwise non-interaction constraints.** A basic constraint in this category is for every antitag not to hybridize to non-complementary tags [3]. For a DNA string $x$ and a set of tags $\mathcal{T}$, let $N_{\mathcal{T}}(x)$ denote the number of tags in $\mathcal{T}$ that contain $x$ as a substring. Using the $c$-token hybridization model, the antitag-to-tag hybridization constraint is formalized as follows:

**(C)** For every feasible tag set $\mathcal{T}$, $N_{\mathcal{T}}(x) \leq 1$ for every DNA string $x$ of weight $c$ or more.

In many assays based on universal tag arrays it is also required to prevent antitag-to-antitag hybridization, since the formation of such antitag-to-antitag duplexes or antitag hair-pin structures prevents reporter probes from performing their function in the solution-based hybridization steps [6, 14]. The combined constraints on antitag hybridization are formalized as follows

**($\bar{C}$)** For every feasible tag set $\mathcal{T}$, $N_{\mathcal{T}}(x) + N_{\mathcal{T}}(\bar{x}) \leq 1$ for every DNA string $x$ of weight $c$ or more.

In the following we use the parameter $\beta \in \{C, \bar{C}\}$ to specify the type of pairwise hybridization constraints.

**Substring occurrences within a tag.** Previous works on DNA tag set design [3, 14] have imposed the following *c-token uniqueness constraint* in addition to constraints $(C)$ and $(\bar{C})$: a DNA string of weight $c$ or more can appear as a substring of a feasible tag at most once. This uniqueness constraint has been added purely for ease of analysis (e.g., it is the key property enabling the DeBruijn sequence based heuristics in [3]), and is *not* required for ensuring correct assay functionality. To our knowledge, no previous work has assessed the impact that adding this constraint has on tag set size. In the following we will use the parameter $\gamma \in \{1, multiple\}$ to specify whether or not the $c$-token uniqueness constraint is enforced.

For every $\alpha \in \{l, h\}$, $\beta \in \{C, \bar{C}\}$, and $\gamma \in \{1, multiple\}$, the maximum tag set design problem with constraints $\alpha, \beta, \gamma$, denoted MTSDP($\alpha|\beta|\gamma$), is the following: given constants $c$ and $l/h$, find a tag set of maximum cardinality satisfying the constraints.

**Previous work on tag set design.** Ben-Dor et al. [3] formalized the $c$-token model for oligonucleotide hybridization and studied the MTSDP($h|C|1$) problem. They established a constructive upperbound on the optimal number of tags for this formulation, and gave a nearly optimal tag selection algorithm based on DeBruijn sequences. Similar upper bounds are established for the MTSDP($l|C|1$) and MTSDP($*|C|1$) problems in [14], which also extends a simple alphabetic tree search strategy originally proposed in [13] to handle all considered problem variants.

For a comprehensive survey of hybridization models, results on the associated formulations for the tag set design problem, and further motivating applications in the area of DNA computing, we direct the reader to [6].

## 3  Integer Linear Programming Formulations for MTSDP($*|C|1$)

Before stating our integer linear program formulation, we introduce some additional notations.

Following [3], a DNA string $x$ of weight $c$ or more is called a *c-token* if all its proper suffixes have weight strictly less than $c$. Clearly, it suffices to enforce constraint $(C)$ for all $c$-tokens $x$. Let $N$ denote the number of $c$-tokens, and $\mathcal{C} = \{c_1, \ldots, c_N\}$ denote the set of all $c$-tokens. The results in [3] imply that $N = \Theta((1+\sqrt{3})^c)$. Note that the weight of a $c$-token can be either $c$ or $c+1$, the latter case being possible only if the $c$-token starts with a strong base (G or C). We let $\mathcal{C}_0 \subseteq \mathcal{C}$ denote the set of $c$-tokens of weight $c+1$ that end with a weak base, i.e., $c$-tokens of the form S$<c-2>$W, where W (S) denote a weak (strong) base, and $<c-2>$ denotes an arbitrary string of weight $c-2$. We also let $\mathcal{C}_2 \subseteq \mathcal{C}$ denote the set of $c$-tokens of weight $c$ that end with a strong base, i.e., $c$-tokens of the form $<c-2>$S.

Clearly, there is at most one $c$-token ending at every letter of a tag. It is easy to see that each $c$-token $x \in \mathcal{C}_0$ contains a proper prefix which is itself a $c$-token, and therefore $x$ cannot be the first

$c$-token of a tag, i.e., cannot be the $c$-token with the leftmost ending. All other $c$-tokens can appear as first $c$-tokens. When a $c$-token in $C \setminus (\mathcal{C}_0 \cup \mathcal{C}_2)$ is the first in a tag, then it must be a prefix of the tag. On the other hand, tokens in $\mathcal{C}_2$ can be the first both in tags that they prefix and in tags in which they are preceded by a weak base not covered by any $c$-token.

The ILP formulation for MTSDP($l|C|1$) uses an auxiliary directed graph $G = (V, E)$ with $V = \{s, t\} \cup \bigcup_{1 \leq i \leq N} V_i$, where $V_i = \{v_i^k \mid |c_i| \leq k \leq l\}$. $G$ has a directed arc from $v_i^k$ to $v_j^{k+1}$ for every triple $i, j, k$ such that $|c_i| \leq k \leq l - 1$ and $c_j$ is obtained from $c_i$ by appending a single nucleotide and removing the maximal prefix that still leaves a valid $c$-token. Finally, $G$ has an arc from $s$ to every $v \in V_{first}$, where $V_{first} = \{v_i^{|c_i|} \mid c_i \in \mathcal{C} \setminus \mathcal{C}_0\} \cup \{v_i^{|c_i|+1} \mid c_i \in \mathcal{C}_2\}$, and an arc from $v_i^l$ to $t$ for every $1 \leq i \leq N$.

We claim that, for $c \leq l$, MTSDP($l|C|1$) can be reformulated as the problem of finding the maximum number of $s$-$t$ paths in $G$ that collectively visit at most one vertex $v_i^k$ for every $i$. Indeed, let $P$ be an $s$-$t$ path and $v_i^k$ be the vertex following $s$ in $P$. If $k = |c_i|$, we associate to $P$ the tag obtained by concatenating $c_i$ with the last letters of the $c$-tokens corresponding to the subsequently visited vertices, until reaching $t$. Otherwise, if $k = |c_i| + 1$ (which implies that $c_i \in \mathcal{C}_2$) we associate to $P$ the two tags obtained by concatenating either A or T with $c_i$ and the last letters of subsequently visited $c$-tokens. The claim follows by observing that at most one of the tags associated with each path can be used in a feasible solution.

Our ILP formulation can be viewed as a generalized version of the integer maximum flow problem in which unit capacity constraints are imposed on *sets of vertices* of $G$ instead of individual vertices. The formulation uses 0/1 variables $x_v$ and $y_e$ for every every vertex $v \in V \setminus \{s, t\}$, respectively arc $e \in E$. These variables are set to 1 if the corresponding vertex or arc is visited by an $s$-$t$ path corresponding to a selected tag. Let $in(v)$ and $out(v)$ denote the set of arcs entering, respectively leaving vertex $v$. The integer program can then be written as follows:

$$\text{maximize} \sum_{v \in V_{first}} x_v \tag{1}$$

subject to

$$x_v = \sum_{e \in in(v)} y_e = \sum_{e \in out(v)} y_e, \quad v \in V \setminus \{s, t\} \tag{2}$$

$$\sum_{v \in V_i} x_v \leq 1, \quad 1 \leq i \leq N \tag{3}$$

$$x_v, y_e \in \{0, 1\}, \quad v \in V \setminus \{s, t\}, e \in E \tag{4}$$

Constraints (2) ensure that variables $y_e$ set to 1 correspond to a set of $s$-$t$ paths, and that a variable $x_v$ is set to 1 if and only if one of these paths passes through $v$.[1] Antitag-to-tag hybridization constraints ($C$) and $c$-token uniqueness are enforced by (3). Finally, the objective (1) corresponds to maximizing the number of selected tags, since the shortest prefix of a tag that is a $c$-token must belong to $\mathcal{C} \setminus \mathcal{C}_0$.

For a token $c_i = c_j X \in \mathcal{C}_0$, where $X \in \{A, T\}$, let $\widehat{c}_i = c_j \bar{a}$. Since both $c_i$ and $\widehat{c}_i$ contain $c_j$ as a prefix, and $c_j$ can appear at most once in a feasible tag set $\mathcal{T}$, it follows that at most one of them can appear in $\mathcal{T}$. Therefore, the following valid inequality can be added to the the ILP formulation (1)–(4) to improve its integrality gap (i.e., the gap between the value of the optimum integer solution and that of the optimal fractional relaxation):

$$\sum_{v \in V_i \cup V_j} x_v \leq 1, \quad c_i \in \mathcal{C}_0, \ c_j = \widehat{c}_i, \ i < j \tag{5}$$

---

[1] Variables $x_v$ can be eliminated by replacing them with the corresponding sums of $x_e$'s; we use them here merely for improving readability. ILP sizes reported in Section 5 refer to the equivalent reduced formulations obtained by eliminating these variables.

The formulation of MTSDP($l|C|1$) has exactly the same objective and constraints for a slightly different graph $G$. Let us define the *tail weight* of a $c$-token $C$, denoted $tail(C)$, as the weight of $C$'s last letter. Also, let $h_i = h$ if $c_i$ has a tail weight of 1 and $h_i = h + 1$ if $c_i$ has a tail weight of 2. We will require that every tag ending with token $c_i$ has total weight of at most $h_i$; it is easy to see that this constraint is not affecting the size of the optimum tag set. We now define the graph $G = (V, E)$ with $V = \{s, t\} \cup \bigcup_{1 \leq i \leq N} V_i$, where $V_i = \{v_i^k \mid w(c_i) \leq k \leq h_i\}$. $G$ contains a directed arc from $v_i^k$ to $v_j^{k+tail(i)}$ for every triple $i, j, k$ such that $|c_i| \leq k \leq h_i - tail(c_i)$ and $c_j$ is obtained from $c_i$ by appending a single nucleotide and removing the maximal prefix that still leaves a valid $c$-token. Finally, $G$ contains arcs from $s$ to every $v \in V_{first}$, where $V_{first}$ is now equal to $\{v_i^{w(c_i)} \mid c_i \in \mathcal{C} \setminus \mathcal{C}_0\} \cup \{v_i^{w(c_i)+1} \mid c_i \in \mathcal{C}_2\}$, plus arcs from every $v_i^k$ to $t$ for every $1 \leq i \leq N$ and $h_i - tail(c_i) < k \leq h_i$.

## 4  Algorithms for MTSDP($*|*|multiple$)

In the following we describe two algorithms for MTSDP($l|C|multiple$); both algorithms can be easily adjusted to handle the other MTSDP($*|*|multiple$) variants. The first algorithm (see Figure 1 for a detailed pseudocode) is similar to the alphabetic tree search algorithms proposed for MTSDP($l|C|1$) in [14]. The algorithm performs an alphabetical traversal of a 4-ary tree representing all $4^l$ possible tags, skipping over subtrees rooted at internal vertices that correspond to tag prefixes including unavailable $c$-tokens. The difference from the MTSDP($l|C|1$) algorithm in [14] lies in the strategy used to mark $c$-tokens as unavailable. While the algorithm in [14] marks a $c$-token $C$ as unavailable as soon as it incorporates it in the current tag prefix (changing $C$'s status back to "available" when forced to backtrack past $C$'s tail), the algorithm in Figure 1 marks a $c$-token as unavailable only when a complete tag is found.

Note that the alphabetic tree search algorithm produces a *maximal* feasible set of tags $\mathcal{T}$, i.e., there is no tag $t$ such that $\mathcal{T} \cup \{t\}$ remains feasible for MTSDP($l|C|multiple$). Hence, every tag of an optimal solution must share at least one $c$-token with tags in $\mathcal{T}$. Since every tag of $\mathcal{T}$ has at most $l - c/2 + 1$ $c$-tokens, it follows that the alphabetic tree algorithm (and indeed, every algorithm that produces a maximal feasible set of tags) has an approximation factor of $l - c/2 + 1$.

We call a tag $t$ *periodic* if $t$ is the length $l$ prefix of an infinite string $x^\infty$, where $x$ is a DNA string with $|x| < |t|$. (Note that a periodic tag $t$ is not necessarily the concatenation of an integer number of copies of its period $x$ as in the standard definition of string periodicity [12].)

The following lemma shows that tag set design algorithms can restrict the search to two simple classes of tags.

**Lemma 1.** *For every $c$ and $l$, there exists an optimal tag set $\mathcal{T}$ in which every tag has the uniqueness property or is periodic.*[2]

*Proof.* Let $\mathcal{T}$ be an optimal tag set. Assume that $\mathcal{T}$ contains a tag $t$ that does not have the uniqueness property, and let $c_{i_1}, \ldots, c_{i_k}$ be the sequence of $c$-tokens occurring in $t$, in left to right order. Since $t$ does not have the uniqueness property, there exist indices $1 \leq j < j' \leq i_k$ such that $c_{i_j} = c_{i_{j'}}$. Let $t'$ be the tag formed by taking the first $l$ letters of the infinite string with $c$-token sequence $(c_{i_j}, \ldots, c_{i_{j'-1}})^\infty$; note that $t'$ is a periodic tag. Since $c$-tokens $c_{i_j}, \ldots, c_{i_{j'-1}}$ do not appear in the tags of $\mathcal{T} \setminus \{t\}$, it follows that $(\mathcal{T} \setminus \{t\}) \cup \{t'\}$ is also optimal. Repeated application of this operation yields the lemma. □

Note that a periodic tag whose shortest period has length $p$ contains as substrings exactly $p$ $c$-tokens, while tags with the uniqueness property contain between $l - c + 1$ and $l - c/2 + 1$ $c$-tokens. Therefore, of the two classes of tags in Lemma 1, periodic tags (particularly those with short periods) make better use of the limited number of available $c$-tokens.

---
[2] Note that the two classes of tags are not disjoint, since there exist periodic tags that have the uniqueness property.

```
Input: Positive integers c and l, c ≤ l
Output: Feasible MTSDP(l|C|multiple) solution 𝒯
─────────────────────────────────────────────────────────────
Mark all c-tokens as available
For every i ∈ {1, 2, ..., l}, B_i ← A
𝒯 ← ∅; Finished ← 0; pos ← 1
While Finished = 0 do
      While the weight of B_1 B_2 ... B_pos < c do
            pos ← pos + 1
      EndWhile
      If the c-token ending B_1 B_2 ... B_pos is available then
            If pos = l then
                  𝒯 ← 𝒯 ∪ {B_1 B_2 ... B_l}
                  Mark all the c-tokens of B_1 B_2 ... B_l as unavailable
                  pos ← [the position where the first c-token of B_1 B_2 ... B_l ends]
                  I ← {i | 1 ≤ i ≤ pos, B_i ≠ G}
                  If I = ∅ then
                        Finished ← 1
                  Else
                        pos ← max{I}
                        B_i ← A for all i ∈ {pos + 1, ..., l}
                        B_pos ← nextbase(B_pos)
                  EndIf
            Else
                  pos ← pos + 1
            EndIf
      Else
            I ← {i | 1 ≤ i ≤ pos, B_i ≠ G}
            If I = ∅ then
                  Finished ← 1
            Else
                  pos ← max{I}
                  B_i ← A for all i ∈ {pos + 1, ..., l}
                  B_pos ← nextbase(B_pos)
            EndIf
      EndIf
EndWhile
```

**Fig. 1.** The alphabetic tree search algorithm for MTSDP($l|C|multiple$). The $nextbase(\cdot)$ function is defined by $nextbase(A) = T$, $nextbase(T) = C$, and $nextbase(C) = G$.

Each periodic tag corresponds to a directed cycle in the graph $H_c$ which has $\mathcal{C}$ as its vertex set, and in which a token $c_i$ is connected by an arc to token $c_j$ iff $c_i$ and $c_j$ can appear consecutively in a tag, i.e., iff $c_j$ is obtained from $c_i$ by appending a single nucleotide and removing the maximal prefix that still leaves a valid $c$-token. Clearly, a vertex-disjoint packing of $n$ cycles in $H_c$ yields a feasible solution for MTSDP($l|C|multiple$) consisting of $n$ tags, since we can extract at least one tag of length $l$ from each cycle, and tags extracted from different cycles do not have common $c$-tokens. This motivates the following:

MAXIMUM VERTEX-DISJOINT DIRECTED CYCLE PACKING Problem: Given a directed graph $G$, find a maximum number of vertex-disjoint directed cycles in $G$.

The next theorem shows that MAXIMUM VERTEX-DISJOINT DIRECTED CYCLE PACKING in arbitrary graphs is unlikely to admit a polynomial approximation scheme. A stronger inapproximability results was recently established for arbitrary graphs by Salavatipour and Verstraete [16], who proved that there is no $O(\log^{1-\varepsilon} n)$-approximation for MAXIMUM VERTEX-DISJOINT DIRECTED CYCLE PACKING unless $NP \subseteq DTIME(2^{polylog n})$. On the positive side, Salavatipour and Verstraete showed that MAXIMUM VERTEX-DISJOINT DIRECTED CYCLE PACKING can be approximated within a factor of $O(\sqrt{n})$ via linear programming techniques, matching the best approximation factor known for the edge-disjoint version of the problem [11].

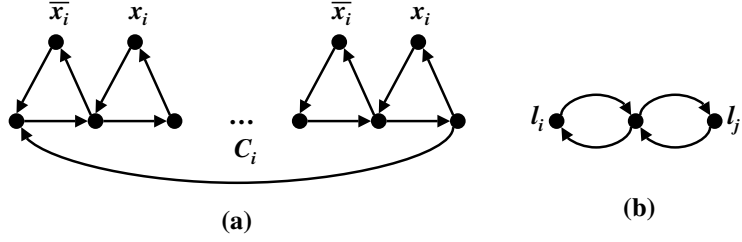

**Fig. 2.** Vertices and arcs added to $G(\phi)$ for (a) variable $x_i$, and (b) clause $l_i \vee l_j$.

**Theorem 1.** MAXIMUM VERTEX-DISJOINT DIRECTED CYCLE PACKING *is APX-hard even for regular directed graphs with in-degree and out-degree of 2.*

*Proof.* We use a reduction from the MAX-2-SAT-3 problem, similar to the one in [8]. An instance $\phi$ of MAX-2-SAT-3 consists of a set $\{c_1, \ldots, c_m\}$ of disjunctive clauses over a ground set $\{x_1, \ldots, x_n\}$ of variables. Each clause consists of at most 2 literals (variables or negations of variable), and each variable appears in at most 3 clauses, counting both negated and non-negated occurrences. The objective is to find a truth assignment that satisfies as many of the clauses as possible. It is known that MAX-2-SAT-3 is APX-hard [2, 5].

Let $m_i$ denote the number of occurrences of variable $x_i$ in a given instance of MAX-2-SAT-3. We construct, in polynomial time, a directed graph $G(\phi)$ as follows. For each variable $x_i$ we add to $G$ a directed cycle $C_i$ of length $4m_i$, plus $2m_i$ additional vertices alternatively labeled by $x_i$ and $\bar{x}_i$, used to close a directed cycle of length 3 with each arc of $C_i$, as in Figure 2(a). For each unary clause we pick a distinct vertex labeled by the *negation* of the respective literal and attach a loop to it. Finally, for each 2-literal clause $c$ we pick 2 vertices labeled by the negations of the literals of $c$, again without reusing labeled vertices between clauses, and use a new vertex to connect them via two length-2 cycles as in Figure 2(b). Note that, for every $i$, at least $2\sum_{i=1}^{n} m_i$ of the labeled vertices remain incident to a single cycle; we will refer to these as "free" labeled vertices.

We claim that every truth assignment that makes $k$ clauses of $\phi$ true can be converted in polynomial time into a set of $k + 2\sum_{i=1}^{n} m_i$ vertex disjoint cycles of $G(\phi)$, and vice-versa. Indeed, for a given truth assignment, select (1) the $2m_i$ length-3 cycles passing through nodes labeled by $\bar{x}_i$ for every variable $x_i$ that is set to true, (2) the $2m_i$ length-3 cycles passing through nodes labeled by $\bar{x}_i$ for every variable $x_i$ that is set to false, and (3) the loop or length-2 cycle passing through a labeled node corresponding to a *false* literal. It is easy to verify that these cycles are vertex-disjoint.

Conversely, let $\mathcal{C}$ be a set of $k + 2\sum_{i=1}^{n} m_i$ vertex disjoint cycles of $G(\phi)$. If any of the cycles $C_i$ is in $\mathcal{C}$, we replace it by the length-3 cycle passing through a free labeled vertex. Similarly, if any of the cycles in $\mathcal{C}$ visits two of the arcs of a 3-cycle (or one of the arcs of a 2-cycle), we replace it by the 3-cycle (respectively 2-cycle) itself. After this transformation we have a set of $k + 2\sum_{i=1}^{n} m_i$ vertex-disjoint loops, 2-cycles, and 3-cycles. We say that a set of cycles is *consistent* if only one of the labels $x_i, \bar{x}_i$ appear in $\mathcal{C}$ for every $i$. If $\mathcal{C}$ is consistent, we choose a truth assignment that makes all literals corresponding to labels in $\mathcal{C}$ true. It is easy to see that at least $k$ of the cycles in $\mathcal{C}$ must be loops and 2-cycles, and clauses corresponding to these cycles are satisfied by the above truth assignment.

Otherwise, we make $\mathcal{C}$ consistent by repeating the following transformation. Let $i$ be an index for which both $x_i$ and $\bar{x}_i$ appear in $\mathcal{C}$. Without loss of generality, assume that $x_i$ appears in only one clause of $\phi$ (recall that, together, $x_i$ and $\bar{x}_i$ can appear in at most 3 clauses). It follows that there is a single loop or 2-cycle $C \in \mathcal{C}$ visiting a vertex labeled by $\bar{x}_i$ – all other vertices labeled by $\bar{x}_i$ are free. Since the $x_i$'s and $\bar{x}_i$'s alternate around $C_i$, the cycles going through vertices labeled by $\bar{x}_i$ can be replaced by at least the same number of 3-cycles going through vertices labeled by $x_i$.

**Table 1.** ILP results for MTSDP($l|C|1$), i.e., tag set design with specified tag length $l$, antitag-to-tag hybridization constraints, and a unique copy of each $c$-token allowed in a tag.

| $l$ | $c$ | #tags [14] | #tags ILP | Upper Bounds LP | Upper Bounds [14] | #constr | #vars | #non-zero | LP time | ILP time |
|---|---|---|---|---|---|---|---|---|---|---|
| 10 | 4 | 7 | 8 | 8.57 | 9 | 406 | 1878 | 6004 | 0.13 | 0.71 |
| 10 | 5 | 23 | 28 | 28.00 | 29 | 1008 | 4600 | 14596 | 2.27 | 5.85 |
| 10 | 6 | 67 | 85 | 85.60 | 96 | 2434 | 10940 | 34470 | 11.40 | 98.25 |
| 10 | 7 | 196 | 259 | 259.67 | 328 | 5808 | 25422 | 79274 | 86.70 | 586.67 |
| 10 | 8 | 655 | – | 853.33 | 1194 | 13554 | 57138 | 175492 | 552.74 | – |
| 20 | 4 | 3 | 3 | 3.53 | 3 | 926 | 4638 | 15244 | 1.05 | 58.46 |
| 20 | 5 | 9 | 10 | 10.50 | 11 | 2448 | 12240 | 40076 | 13.72 | 381.33 |
| 20 | 6 | 26 | 29 | 29.87 | 32 | 6354 | 31860 | 104270 | 182.96 | 12448.61 |
| 20 | 7 | 75 | – | 88.00 | 93 | 16528 | 82662 | 270194 | 2675.68 | – |
| 20 | 8 | 213 | – | 257.23 | 275 | 42834 | 213578 | 697292 | 134525.81 | – |

**Table 2.** ILP results for MTSDP($h|C|1$), i.e., tag set design with specified minimum tag weight $h$, antitag-to-tag hybridization constraints, and a unique copy of each $c$-token allowed in a tag.

| $h$ | $c$ | #tags [14] | #tags ILP | Upper Bounds LP | Upper Bounds [3] | #constr | #vars | #non-zero | LP time | ILP time |
|---|---|---|---|---|---|---|---|---|---|---|
| 15 | 4 | 6 | 7 | 7.00 | 7 | 610 | 2966 | 9612 | 0.45 | 9.04 |
| 15 | 5 | 18 | 21 | 21.09 | 21 | 1550 | 7456 | 23998 | 5.66 | 117.62 |
| 15 | 6 | 47 | 63 | 63.20 | 63 | 3830 | 18322 | 58752 | 54.43 | 2665.39 |
| 15 | 7 | 149 | 192 | 192.00 | 192 | 9406 | 44416 | 141638 | 544.95 | 3644.85 |
| 15 | 8 | 460 | – | 588.00 | 590 | 22766 | 105746 | 334904 | 7153.87 | – |
| 28 | 4 | 3 | 3 | 3.30 | 3 | 1286 | 6554 | 21624 | 1.88 | 132.78 |
| 28 | 5 | 8 | 9 | 9.67 | 9 | 3422 | 17388 | 57122 | 34.66 | 1137.21 |
| 28 | 6 | 22 | 27 | 27.48 | 27 | 8926 | 45518 | 149492 | 392.42 | 18987.09 |
| 28 | 7 | 64 | – | 78.55 | 78 | 23342 | 118828 | 389834 | 7711.41 | – |
| 28 | 8 | 175 | – | – | 224 | 60830 | 309118 | 1013244 | – | – |

To complete the proof of the theorem, notice that the optimum number of satisfiable clauses, $k_{opt}$, is at least $m/2$, since we can repeatedly assign a variable such that at least half of the clauses containing it are satisfied. Hence, $\sum_{i=1}^{n} m_i \leq 2m \leq 4k_{opt}$. If there exists a polynomial time algorithm with an approximation factor of $\frac{1}{1-\varepsilon}$ for MAXIMUM VERTEX-DISJOINT DIRECTED CYCLE PACKING, we can run it on $G(\phi)$ to get a set $\mathcal{C}$ of at least $k + 2\sum_{i=1}^{n} m_i \geq \frac{1}{1-\varepsilon}(k_{opt} + 2\sum_{i=1}^{n} m_i)$ vertex disjoint cycles, and then convert $\mathcal{C}$ as above into a truth assignment satisfying $k \geq \frac{1+8\varepsilon}{1-\varepsilon} k_{opt}$ clauses of $\phi$. □

We use a simple greedy algorithm to solve MAXIMUM VERTEX-DISJOINT DIRECTED CYCLE PACKING for the graph $H_c$: we enumerate possible tag periods in pseudo-lexicographic order, and check for each period if all $c$-tokens are available for the resulting tag. We refer to this algorithm as the *greedy cycle packing algorithm*, since it is equivalent to packing cycles greedily in order of length.

## 5 Experimental results

Tables 1 and 2 give ILP statistics (number of constraints, number of variables, and number of non-zero coefficients), LP and ILP runtime, and LP and ILP solution values for MTSDP($l|C|1$) and MTSDP($h|C|1$). We also include the upper bounds established in [14] and [3] for these problems, and the number of tags found by using the alphabetic tree search algorithm in [14]. We solved all integer programs and their fractional relaxations using the CPLEX 9.0 commercial solver with default parameters run using a single CPU on a dual 2.8 GHz Dell PowerEdge 2600 Linux server. Missing entries did not complete in 10 hours.

**Table 3.** Results for MTSDP($*|C|multiple$), i.e., tag set design with antitag-to-tag hybridization constraints and multiple copies of a $c$-token allowed in a tag.

| $l/h$ | $c$ | One $c$-token copy | | Multiple $c$-token copies | | | | |
| --- | --- | --- | --- | --- | --- | --- | --- | --- |
| | | Algorithm in [14] | | Tree search | | Cycle packing + Tree search | | |
| | | tags | $c$-tokens | tags | $c$-tokens | tags | $c$-tokens | % cyclic |
| $l = 20$ | 4 | 3 | 51 | 14 | 59 | 17 | 40 | 100.0 |
| | 5 | 9 | 146 | 31 | 165 | 40 | 140 | 100.0 |
| | 6 | 26 | 404 | 53 | 433 | 72 | 293 | 98.6 |
| | 7 | 75 | 1100 | 124 | 1179 | 178 | 928 | 99.4 |
| | 8 | 213 | 2976 | 281 | 3095 | 383 | 2411 | 97.1 |
| | 9 | 600 | 7931 | 711 | 8230 | 961 | 7102 | 96.9 |
| | 10 | 1667 | 20771 | 1835 | 21400 | 2344 | 19691 | 95.1 |
| $h \geq 28$ | 4 | 3 | 58 | 14 | 61 | 17 | 40 | 100.0 |
| | 5 | 8 | 150 | 32 | 174 | 40 | 140 | 100.0 |
| | 6 | 22 | 398 | 44 | 432 | 72 | 300 | 98.6 |
| | 7 | 64 | 1119 | 118 | 1200 | 178 | 934 | 99.4 |
| | 8 | 175 | 2918 | 239 | 3037 | 379 | 2405 | 96.6 |
| | 9 | 531 | 8431 | 632 | 8622 | 943 | 6969 | 96.5 |
| | 10 | 1428 | 21707 | 1570 | 22145 | 2260 | 19270 | 94.1 |

**Table 4.** Results for MTSDP($*|\bar{C}|multiple$), i.e., tag set design with both antitag-to-tag and antitag-to-antitag hybridization constraints and multiple copies of a $c$-token allowed in a tag.

| $l/h$ | $c$ | One $c$-token copy | | Multiple $c$-token copies | | | | |
| --- | --- | --- | --- | --- | --- | --- | --- | --- |
| | | Algorithm in [14] | | Tree search | | Cycle packing + Tree search | | |
| | | tags | $c$-tokens | tags | $c$-tokens | tags | $c$-tokens | % cyclic |
| $l = 20$ | 4 | 1 | 17 | 10 | 35 | 10 | 25 | 100.0 |
| | 5 | 4 | 65 | 17 | 83 | 23 | 85 | 100.0 |
| | 6 | 13 | 200 | 30 | 241 | 41 | 171 | 97.6 |
| | 7 | 37 | 537 | 68 | 585 | 97 | 512 | 99.0 |
| | 8 | 107 | 1480 | 147 | 1619 | 202 | 1268 | 98.0 |
| | 9 | 300 | 3939 | 362 | 4124 | 512 | 3799 | 96.3 |
| | 10 | 844 | 10411 | 934 | 10869 | 1204 | 10089 | 95.8 |
| $h \geq 28$ | 4 | 1 | 22 | 10 | 36 | 10 | 25 | 100.0 |
| | 5 | 4 | 74 | 17 | 84 | 23 | 85 | 100.0 |
| | 6 | 12 | 213 | 29 | 238 | 41 | 178 | 97.6 |
| | 7 | 32 | 559 | 64 | 586 | 97 | 518 | 99.0 |
| | 8 | 90 | 1489 | 135 | 1632 | 199 | 1238 | 98.0 |
| | 9 | 263 | 4158 | 329 | 4314 | 504 | 3760 | 95.8 |
| | 10 | 714 | 10837 | 809 | 11250 | 1163 | 9937 | 93.6 |

The ILP solutions can be found in practical time for small values of $c$, which are appropriate for universal tag array applications, such as the emerging microfluidics-based labs-on-a-chip, where moderate multiplexing rates are sufficient and ensuring high hybridization stringency is costly. For all cases where the optimum could be computed, the difference between the optimal fractional and integer solution values was smaller than 1, indicating why CPLEX can solve to optimality these ILPs despite their size. Furthermore, ILP results confirm the extremely high quality of the upperbound established for MTSDP($h|C|1$) in [3]; the upperbound established in [14] for MTSDP($l|C|1$) appears to be somehow weaker.

Tables 3 and 4 give the results obtained for MTSDP($*|*|multiple$) by the alphabetic tree search algorithm in Figure 1 respectively by the greedy cycle packing algorithm (in our implementation, we impose an upper bound of 15 on the length of the cycles that we try to pack) followed by running the alphabetic tree search algorithm with the $c$-tokens occurring in the selected cycles already marked as unavailable. Performing cycle packing significantly improves the results compared to running the alphabetic tree search algorithm alone; as shown in the tables, most of the resulting tags are found in the cycle packing phase of the combined algorithm.

Across all instances, the combined algorithm increases the number of tags by at least 40% compared to the MTSDP($*|*|1$) algorithm in [14]; the improvement is much higher for smaller values of $c$. Quite notably, although the number of tags is increased, the tag sets found by the combined algorithm use a *smaller* total number of $c$-tokens. Thus, these tag sets are less likely to cross-hybridize to the primers used in the reporter probes, enabling higher tag utilization rates during tag assignment [4, 14].

## 6 Conclusions

In this paper we proposed new solution methods for designing tag sets for universal DNA arrays. We have shown that optimal solutions can be found in practical time for moderate problem sizes by using integer linear programming, and that the use of periodic tags leads to increases of over 40% in the number of tags, with simultaneous increases in effective tag utilization rates during tag assignment. Our algorithms use simple greedy strategies, and can be easily modified to incorporate additional practical design constraints, such as preventing the formation of hairpin secondary structures, or disallowing specific nucleotide sequences such as runs of 4 identical nucleotides [13].

An interesting open problem is to find tight upper bounds and exact methods for the MTSDP($*|*|multiple$) formulations. Settling the approximation complexity of MAXIMUM VERTEX-DISJOINT DIRECTED CYCLE PACKING is another interesting problem.

## References

1. Affymetrix, Inc. Geneflex tag array technical note no. 1, available online at http://www.affymetrix.com/support/technical/technotes/genflex_technote.pdf.
2. G. Ausiello, M. Protasi, A. Marchetti-Spaccamela, G. Gambosi, P. Crescenzi, and V. Kann. *Complexity and Approximation: Combinatorial Optimization Problems and Their Approximability Properties*. Springer-Verlag New York, Inc., 1999.
3. A. Ben-Dor, R. Karp, B. Schwikowski, and Z. Yakhini. Universal DNA tag systems: a combinatorial design scheme. *Journal of Computational Biology*, 7(3-4):503–519, 2000.
4. A. BenDor, T. Hartman, B. Schwikowski, R. Sharan, and Z. Yakhini. Towards optimally multiplexed applications of universal DNA tag systems. In *Proc. 7th Annual International Conference on Research in Computational Molecular Biology*, pages 48–56, 2003.
5. P. Berman and M. Karpinski. On some tighter inapproximability results. In *Proc. 26th Intl. Colloquium on Automata, Languages and Programming*, pages 200–209, 1999.
6. A. Brenneman and A. Condon. Strand design for biomolecular computation. *Theor. Comput. Sci.*, 287(1):39–58, 2002.
7. S. Brenner. Methods for sorting polynucleotides using oligonucleotide tags. *US Patent 5,604,097*, 1997.
8. A. Caprara, A. Panconesi, and R. Rizzi. Packing cycles in undirected graphs. *Journal of Algorithms*, 48(1):239–256, 2003.
9. J.N. Hirschhorn et al. SBE-TAGS: An array-based method for efficient single-nucleotide polymorphism genotyping. *PNAS*, 97(22):12164–12169, 2000.
10. L. Kaderali. *Selecting Target Specific Probes for DNA Arrays*. PhD thesis, Köln University, 2001.
11. M. Krivelevich, Z. Nutov, and R. Yuster. Approximation algorithms for cycle packing problems. In *Proc. ACM-SIAM Annual Symposium on Discrete Algorithms*, pages 556–561, 2005.
12. M. Lothaire. *Combinatorics on Words*, volume 17 of *Encylopedia of Mathematics and Its Applications*, pages xix+238. Addison-Wesley, 1983.
13. M.S. Morris, D.D. Shoemaker, R.W. Davis, and M.P. Mittmann. Selecting tag nucleic acids. *U.S. Patent 6,458,530 B1*, 2002.
14. I.I. Măndoiu, C. Prăjescu, and D. Trincă. Improved tag set design and multiplexing algorithms for universal arrays. In *Proc. 1st International Workshop on Bioinformatics Research and Applications (IWBRA)*, 2005 (to appear).
15. N.P. Gerry et al. Universal DNA microarray method for multiplex detection of low abundance point mutations. *J. Mol. Biol.*, 292(2):251–262, 1999.

16. M.R. Salavatipour and J. Verstraete. Disjoint cycles: Integrality gap, hardness, and approximation. In *Proc. 11th Conference on Integer Programming and Combinatorial Optimization (IPCO)*, 2005 (to appear).
17. J. SantaLucia. A unified view of polymer, dumbbell, and oligonucleotide DNA nearest-neighbor thermodynamics. *Proc. Natl. Acad. Sci. USA*, 95:1460–1465, 1998.
18. R.B. Wallace, J. Shaffer, R.F. Murphy, J. Bonner, T. Hirose, and K. Itakura. Hybridization of synthetic oligodeoxyribonucleotides to phi chi 174 DNA: the effect of single base pair mismatch. *Nucleic Acids Res.*, 6(11):6353–6357, 1979.